\documentclass[journal]{IEEEtran}

\usepackage{amsmath,amssymb}
\usepackage{graphicx}
\usepackage{cite}

\graphicspath{{figures/}}

\title{ \huge Joint Localization and Data Detection in Ambient Backscatter Communications: Uncertainty-Preserving Inference and Cross-Frame Geometry Consensus}

\author{
	Xianhua Yu, Dong Li,~\IEEEmembership{Senior Member,~IEEE}, Bowen Gu, Tianhao Liang and Tingting Zhang
	\thanks{Xianhua Yu is with the School of Electrical Engineering and Intelligentization, Dongguan University of Technology, Dongguan, China (e-mail: xianhuacn@foxmail.com). Dong Li is with the School of Computer Science and Engineering, Macau University of Science and Technology, Macau 999078, China (e-mail: dli@must.edu.mo). Bowen Gu is with the School of Computer Science and Technology, Xinjiang University, Urumqi, Xinjiang 830049, China, and also with Xinjiang Multimodal Intelligent Processing and Information Security Engineering Technology Research Center, Urumqi, Xinjiang 830049, China (e-mail: bwgu@xju.edu.cn). Tianhao Liang and Tingting Zhang are with the School of Information Science and Technology and the Guangdong Provincial Key Laboratory of Space-Aerial Networking and Intelligent Sensing, Harbin Institute of Technology, Shenzhen (e-mails: liangth@hit.edu.cn, zhangtt@hit.edu.cn).
	}
	
}

\begin{document}

\maketitle
\pagestyle{empty}
\thispagestyle{empty}

\begin{abstract}
	We address joint continuous localization and data detection in ambient backscatter communication, where weak observations can leave competing geometry hypotheses plausible and a staged point-estimate interface can discard information about this ambiguity. We develop the \emph{Geometry-aware Frame Network (GeoFrameNet)}, which projects channel frequency responses onto a physics-derived delay--angle-of-arrival lattice and aggregates sign-robust evidence across orthogonal frequency-division multiplexing symbols to form a shared geometry posterior over physically feasible candidates. Localization uses the full posterior with candidate-specific refinement; detection combines candidate-conditioned differential logits using posterior weights renormalized over selected candidates. Bit supervision guides geometry scoring. For a fixed device, the \emph{Cross-Frame Geometry Consensus Network (CFGC-Net)} fuses frozen GeoFrameNet candidate logits and features across frames for localization while preserving frame-wise detection outputs. On an independent test set, GeoFrameNet achieves lower localization root-mean-square error (RMSE) than an all-symbol multiple-measurement-vector sparse Bayesian learning (MMV-SBL) baseline at all eight evaluated signal-to-noise ratio (SNR) points, with simultaneous bit error rate and RMSE reductions from $-30$ to $-22.5$~dB. At $-30$~dB SNR, the respective RMSEs are 3.1573 and 18.8859~m. CFGC-Net achieves the lowest aggregate RMSE among the evaluated fusion strategies for two, four, and eight frames.
\end{abstract}


\section{Introduction}
\label{sec:introduction}

\subsection{Background}

Ambient backscatter communication (AmBC) enables a passive backscatter device (BD) to convey information by modulating and reflecting ambient radio-frequency (RF) signals, thereby eliminating the need for an active RF chain at the BD \cite{Liu2013Ambient,Parks2014Turbocharging,Kellogg2014WiFi}. This operating principle is well suited to batteryless and intermittently powered Internet of Things (IoT) deployments, where device complexity and energy consumption are tightly constrained. Reliable reception, however, remains challenging because the backscattered signal experiences cascaded source--BD and BD--receiver attenuation and is typically very weak \cite{Huynh2018Survey,Alevizos2023Batteryless,Gu2025Gridlock}.

Beyond connectivity, passive IoT applications may require both the data conveyed by a BD and its physical location \cite{RenLiu2023SLABC,Ouzane2025OTSM,Ma2025Integrated}. Localization provides spatial context for tracking, inventory, and location-aware monitoring, motivating joint recovery of data and position from the same weak backscatter observation.

\subsection{Related Work and Research Gap}

AmBC data detection has been extensively studied under different levels of source and channel knowledge.
Likelihood-based and noncoherent receivers address unknown or partially known ambient excitation and channel coefficients \cite{Wang2016Detection,Qian2017Noncoherent}. Under ambient orthogonal frequency-division multiplexing (OFDM) illumination, receivers exploit cyclic-prefix repetition, spectral structure, null subcarriers, and multi-antenna processing for interference suppression and
symbol recovery \cite{Yang2018Modulation,Darsena2017Modeling, Duan2018MultiAntenna,ElMossallamy2019OFDM}. Learning-based approaches have further been investigated for constellation learning, transfer across channel conditions, and joint channel/signal inference \cite{Zhang2019Constellation,Liu2021Transfer,Zargari2024Deep}. Attention-aided backscatter detection has also been studied under insufficient training data \cite{Yu2024Attention}. Related work estimates the numbers of active and inactive backscatter tags \cite{Yu2024NumberDetection}, a tag-population inference problem distinct from recovering the payload and position of a single BD. In parallel, backscatter localization has exploited multipath and temporal signatures, wideband delay and angle information, dedicated sensing configurations, and frequency hopping \cite{WangKatabi2013PinIt,Wang2013RFCompass, Yang2014Tagoram,Ma2017RFind,Zhang2020Rover, Bae2023Hawkeye,Hou2024LoBaCa}. These detection and localization studies primarily address the two tasks separately, whereas the present work considers
their joint recovery from the same weak ambient-backscatter observation.

Joint localization and communication have also been investigated under different backscatter waveform and sensing architectures \cite{RenLiu2023SLABC,Ouzane2025OTSM,Ma2025Integrated,Ma2025Phaseless,Ma2026FrequencyDiverse,Ren2026ANM,Li2026Integrated}. The ambient-OFDM receiver in \cite{Xu2024JointLocalization} uses a two-stage architecture: delay and angle of arrival (AoA) are estimated from the channel frequency response (CFR) of the first OFDM symbol using orthogonal matching pursuit or off-grid sparse Bayesian learning (SBL). The estimated geometry is then used for symbol-wise least-squares (LS) coefficient estimation and differential detection over the frame. The SBL extension mitigates dictionary discretization error through continuous off-grid refinement, providing a tractable geometry-first decomposition of the coupled inference problem.

The reliability of this geometry estimate depends on how well a finite, noisy CFR distinguishes competing delay--AoA hypotheses. Delay and AoA are encoded jointly in the spatial--frequency phase structure across receive antennas and pilot subcarriers, rather than supplied as separate measurements. Estimation accuracy depends on the available spatial and frequency apertures and the observation
signal-to-noise ratio (SNR), as reflected by the performance reported in \cite{Xu2024JointLocalization}. Under weak observations, the CFR may therefore leave several physically
feasible hypotheses difficult to distinguish. This observation-induced ambiguity constitutes \emph{geometry uncertainty}, which is distinct from the dictionary discretization error addressed by off-grid estimation \cite{Yang2013OffGrid}.

Although SBL performs probabilistic inference within the localization stage, the receiver in
\cite{Xu2024JointLocalization} fixes the resulting delay--AoA estimates when constructing the steering matrices for subsequent LS estimation and differential detection. The remaining OFDM symbols are processed conditionally on these estimates: they do not update the frame-shared geometry, and competing hypotheses no longer contribute to detection. An inaccurate geometry estimate can therefore introduce
steering mismatch and degrade coefficient estimation and bit detection. The limitation under weak observations lies in the point-estimate interface between geometry inference and data detection.

Addressing this limitation requires exploiting repeated observations of the frame-shared geometry while retaining any residual ambiguity for both receiver tasks. Under differential binary phase-shift keying (DBPSK), however, reflection symbols in $\{-1,+1\}$ can reverse the sign of the common complex response across OFDM symbols, so direct coherent accumulation may cancel rather than reinforce geometric evidence. The challenge is therefore to combine these sign-modulated observations to strengthen geometry
inference and allow competing hypotheses to participate in localization and detection whenever the complete frame remains inconclusive.

\subsection{Proposed Approach and Contributions}

To meet these requirements, we develop the \emph{Geometry-aware Frame Network (GeoFrameNet)} for the single-BD ambient-OFDM setting considered here. GeoFrameNet incorporates the known delay--AoA propagation structure following the general principle of model-informed learning \cite{He2019ModelDriven,Shlezinger2023ModelBased}. Each OFDM-symbol CFR is projected onto a physics-derived dense delay--AoA lattice, and sign-robust magnitude- and energy-based evidence is aggregated across the complete frame. In this way, all OFDM symbols can contribute to discrimination among candidate geometries despite the DBPSK-dependent sign reversals identified above. The resulting frame-level evidence is then converted into a learned geometry posterior over the physically feasible candidate set.

This shared posterior serves as the interface from geometry inference to both receiver tasks. Continuous localization retains the full posterior and combines it with candidate-specific delay--AoA refinement to obtain a continuous position estimate. Data detection instead combines candidate-conditioned differential logits using posterior masses renormalized over a compact set of high-probability candidates. Multiple high-probability geometry hypotheses can therefore continue to contribute to both localization and data detection. During training, the bit-detection loss further influences geometry scoring through the posterior-derived weights, while a stop-gradient operation prevents the detector-content path from directly updating the shared candidate representation. The candidate-specific refinement separately addresses local finite-lattice mismatch.

We further consider repeated communication frames in which the same BD remains at a fixed position. Because the geometry is shared across frames whereas the DBPSK payload, composite phase, and receiver noise are frame specific, directly averaging complex CFRs can cause destructive combination, while averaging only final Cartesian position estimates discards candidate-level geometry information before fusion. We therefore develop the Cross-Frame Geometry Consensus Network (CFGC-Net), which fuses frozen GeoFrameNet candidate logits and features before forming the final position estimate while leaving all frame-wise data-detection outputs unchanged.

Accordingly, GeoFrameNet differs from the staged receiver in \cite{Xu2024JointLocalization} in three linked aspects: it constructs geometry evidence from the complete frame using sign-robust statistics, retains a posterior over competing physical hypotheses rather than passing a single geometry
estimate across stages, and uses task-specific posterior inference for continuous localization and differential detection. The main contributions are summarized as follows.
\begin{enumerate}
	\item We develop sign-robust full-frame geometry inference for GeoFrameNet. A physics-derived dense delay--AoA lattice organizes the CFR observations, and magnitude- and energy-based evidence is aggregated across OFDM symbols to strengthen inference of the frame-shared geometry despite
	DBPSK-dependent sign reversals.
	
	\item We develop posterior-mediated joint inference that avoids an intermediate geometry point estimate. Continuous localization uses the full geometry posterior with candidate-specific refinement, whereas data detection combines candidate-conditioned differential logits using
	renormalized posterior masses over a high-probability subset. During training, the bit loss influences geometry scoring through these weights, while a stop-gradient operation
	blocks the detector-content path to the shared candidate representation.
	
	\item We develop CFGC-Net as a candidate-domain localization extension for repeated frames from a fixed BD. It learns a residual correction to the mean geometry logits from averaged candidate features and applies the frozen GeoFrameNet refinement mapping to those features, while preserving all frame-wise data-detection outputs.
	
	\item Extensive simulations show that GeoFrameNet achieves lower localization root-mean-square error (RMSE) than the all-symbol multiple-measurement-vector SBL (MMV-SBL) baseline at all  evaluated SNR points, with simultaneous bit error rate (BER) and RMSE reductions at the weakest points from $-30$ to $-22.5$~dB. Ablations assess full-frame aggregation, communication-aware geometry scoring, multi-hypothesis inference, and continuous refinement. CFGC-Net achieves the lowest aggregate localization RMSE among the evaluated cross-frame strategies for $F\in\{2,4,8\}$.
\end{enumerate}
	
\section{System Model and Problem Formulation}
\label{sec:system_model}

\begin{figure*}[t]
	\centering
	\includegraphics[width=0.77\textwidth]{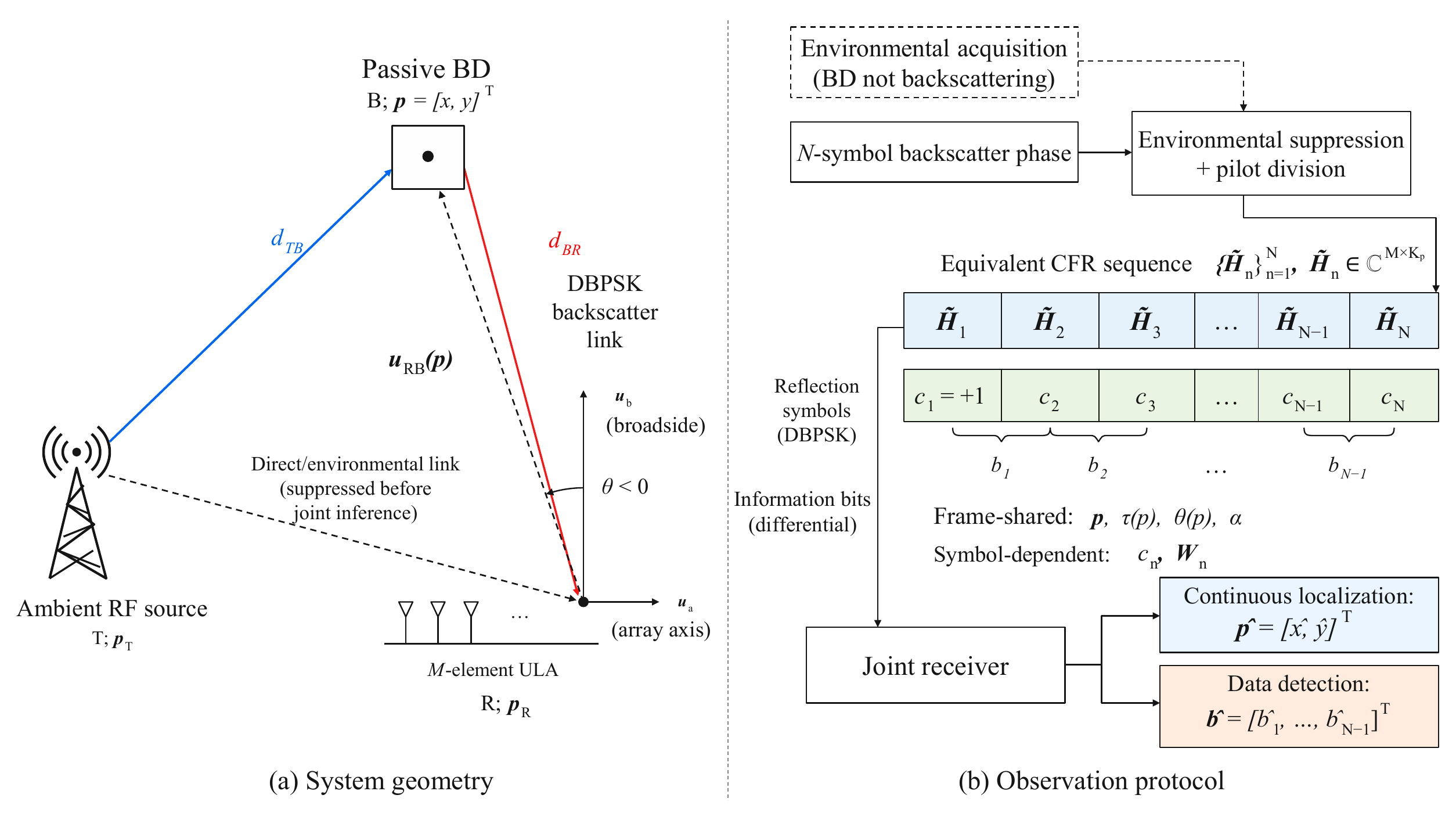}
	\caption{Considered AmBC system and frame structure. The joint receiver operates on the equivalent post-suppression CFR sequence of the subsequent $N$-symbol backscatter phase.}
	\label{fig:system_frame}
\end{figure*}
\subsection{Geometry and Propagation Model}

As illustrated in Fig.~\ref{fig:system_frame}, we consider an ambient backscatter communication system comprising a single-antenna ambient RF source $\mathsf T$, a single-antenna backscatter device (BD) $\mathsf B$, and a receiver $\mathsf R$ equipped with an $M$-element uniform linear array (ULA). The source continuously emits OFDM signals with $K$ subcarriers and subcarrier spacing $\Delta f$. The BD conveys information by modulating its reflection state, while the receiver estimates both the conveyed information and the continuous two-dimensional BD position $\mathbf p=[x,y]^T\in\mathcal P\subset\mathbb R^2$. The known source and receiver positions are denoted by $\mathbf p_{\mathsf T}$ and $\mathbf p_{\mathsf R}$, respectively.

The source-to-BD and BD-to-receiver distances are
\begin{equation}
	\begin{aligned}
		d_{\mathsf{TB}}(\mathbf p)
		&=\|\mathbf p-\mathbf p_{\mathsf T}\|_2,\\
		d_{\mathsf{BR}}(\mathbf p)
		&=\|\mathbf p-\mathbf p_{\mathsf R}\|_2.
	\end{aligned}
	\label{eq:link_distances}
\end{equation}
Under the adopted single-dominant-path backscatter model, the bistatic propagation delay is
\begin{equation}
	\tau(\mathbf p)
	=
	\frac{
		d_{\mathsf{TB}}(\mathbf p)+d_{\mathsf{BR}}(\mathbf p)}
	{c_0},
	\label{eq:bistatic_delay}
\end{equation}
where $c_0$ denotes the propagation speed.
Let
$d_{\mathsf{TR}}
=\|\mathbf p_{\mathsf R}-\mathbf p_{\mathsf T}\|_2$.
By the triangle inequality, every physical BD position satisfies
\[
c_0\tau(\mathbf p)
=
d_{\mathsf{TB}}(\mathbf p)
+
d_{\mathsf{BR}}(\mathbf p)
\geq
d_{\mathsf{TR}}.
\]

Equality occurs when the BD lies on the source--receiver line segment. This boundary is physically feasible but nonunique under the adopted delay--AoA inverse. We therefore focus on the nondegenerate domain $c_0\tau>d_{\mathsf{TR}}$; the corresponding posterior support is constructed in Section~\ref{sec:proposed_receiver}. 

Let $\mathbf u_a\in\mathbb R^2$ denote the unit vector along the known ULA axis and $\mathbf u_b\in\mathbb R^2$ the corresponding positive-broadside unit vector. The AoA is parameterized through the receiver-to-BD look direction
\begin{equation}
	\mathbf u_{\mathsf{RB}}(\mathbf p)
	=
	\frac{\mathbf p-\mathbf p_{\mathsf R}}
	{d_{\mathsf{BR}}(\mathbf p)}.
	\label{eq:receiver_to_bd_direction}
\end{equation}

Restricting the considered region to the positive-broadside half-plane resolves the visible ULA branch:
\begin{equation}
	\begin{aligned}
		\sin\theta(\mathbf p)
		&=
		\mathbf u_a^T\mathbf u_{\mathsf{RB}}(\mathbf p),\\
		\cos\theta(\mathbf p)
		&=
		\mathbf u_b^T\mathbf u_{\mathsf{RB}}(\mathbf p)\geq0,\\
		\theta(\mathbf p)
		&=
		\arcsin\!\left(
		\mathbf u_a^T\mathbf u_{\mathsf{RB}}(\mathbf p)
		\right)
		\in[-\pi/2,\pi/2].
	\end{aligned}
	\label{eq:position_to_aoa}
\end{equation}

With antenna spacing $d_a$ and carrier wavelength $\lambda_c$, the receive-array steering vector is
\begin{equation}
	\mathbf v(\theta)
	=
	\left[
	1,
	e^{j2\pi\frac{d_a}{\lambda_c}\sin\theta},
	\ldots,
	e^{j2\pi(M-1)\frac{d_a}{\lambda_c}\sin\theta}
	\right]^T,
	\label{eq:aoa_steering}
\end{equation}
where the positive spatial-phase convention follows \cite{Xu2024JointLocalization}. Let $\mathcal K_p=\{k_1,\ldots,k_{K_p}\}\subseteq\{0,\ldots,K-1\}$ denote the pilot-subcarrier set. For uniformly spaced pilots, let $D_p$ denote the pilot-subcarrier index spacing. Retaining the actual pilot indices, the delay steering vector is
\begin{equation}
	\mathbf d(\tau)
	=
	\left[
	e^{-j2\pi k_1\Delta f\tau},
	\ldots,
	e^{-j2\pi k_{K_p}\Delta f\tau}
	\right]^T.
	\label{eq:delay_steering}
\end{equation}
Neither AoA nor delay is supplied to the receiver as a separate scalar measurement. Rather, the spatial sine $s=\sin\theta$ is encoded in the inter-element phase progression of $\mathbf v(\theta)$, whereas the delay $\tau$ is encoded in the pilot-frequency phase progression of $\mathbf d(\tau)$.

For later use, define the normalized delay and spatial sine as
\[
\widetilde\tau
=
D_p\Delta f\,\tau,
\quad
s=\sin\theta,
\]
and let
\[
\boldsymbol\Delta_{\mathsf{TR}}
=
\mathbf p_{\mathsf R}-\mathbf p_{\mathsf T},
~
\mathbf u(s)
=
s\mathbf u_a
+
\sqrt{1-s^2}\,\mathbf u_b,
~
L(\widetilde\tau)
=
\frac{c_0\widetilde\tau}{D_p\Delta f}.
\]
Over the nondegenerate domain
$L(\widetilde\tau)>d_{\mathsf{TR}}$, the visible-branch
delay--AoA inverse is
\begin{equation}
	\begin{aligned}
		d_{\mathsf{BR}}(\widetilde\tau,s)
		&=
		\frac{
			L^2(\widetilde\tau)
			-
			\|\boldsymbol\Delta_{\mathsf{TR}}\|_2^2
		}{
			2\left[
			L(\widetilde\tau)
			+
			\boldsymbol\Delta_{\mathsf{TR}}^T\mathbf u(s)
			\right]
		},
		\\
		\boldsymbol\psi(\widetilde\tau,s)
		&=
		\mathbf p_{\mathsf R}
		+
		d_{\mathsf{BR}}(\widetilde\tau,s)\mathbf u(s).
	\end{aligned}
	\label{eq:visible_branch_inverse}
\end{equation}
For $L(\widetilde\tau)>d_{\mathsf{TR}}$, the denominator in \eqref{eq:visible_branch_inverse} is positive, so an exact delay--spatial-sine pair determines a unique position on the selected visible branch. This geometric uniqueness does not imply that a finite noisy CFR uniquely identifies that pair.

Let \(L_{\mathsf{TB}}(\mathbf p)\) and \(L_{\mathsf{BR}}(\mathbf p)\) denote the source--BD and BD--receiver power gains, respectively. The composite backscatter-path coefficient is
\begin{equation}
	\alpha(\mathbf p)
	=
	\sqrt{
		L_{\mathsf{TB}}(\mathbf p)
		L_{\mathsf{BR}}(\mathbf p)
	}
	e^{j\phi},
	\label{eq:backscatter_coefficient}
\end{equation}
where $\phi$ denotes the unknown frame-static composite propagation/reflection phase, and any constant BD reflection-magnitude factor is absorbed into $\alpha(\mathbf p)$. Hence, $|\alpha(\mathbf p)|$ is determined by the geometry-dependent link gains, whereas its phase remains unknown. We suppress the dependence of $\alpha$ on $\mathbf p$ when clear.

\subsection{Frame Structure and Equivalent CFR Observation}

Following the protocol setting in \cite{Xu2024JointLocalization}, the environmental response is acquired during a preceding nonbackscattering phase and suppressed before joint inference. We adopt the resulting equivalent CFR model without a separate residual-interference term. The acquisition-phase observations are used only for suppression; the joint receiver takes the subsequent $N$ post-suppression CFR matrices as its input.

During this phase, the synchronized source continuously emits OFDM symbols, while the BD applies one DBPSK reflection symbol during each OFDM symbol. Let $c_n\in\{-1,+1\}$ denote the reflection symbol in symbol $n$, with the fixed reference $c_1=+1$. If $b_n\in\{0,1\}$ denotes the information bit associated with transition $n\rightarrow n+1$, then
\begin{equation}
	\begin{aligned}
		c_{n+1}
		&=c_n(1-2b_n),\\
		b_n
		&=
		\frac{1-c_{n+1}c_n}{2},
		\quad n=1,\ldots,N-1.
	\end{aligned}
	\label{eq:dbpsk_mapping}
\end{equation}

Hence, the $N$ reflection symbols convey $\mathbf b=[b_1,\ldots,b_{N-1}]^T\in\{0,1\}^{N-1}$.

The designated ambient-source pilot symbols are known at the receiver and have unit modulus. After environmental-link suppression and elementwise division by these pilots, the receiver obtains the equivalent pilot-domain CFR matrix $\widetilde{\mathbf H}_n\in\mathbb C^{M\times K_p}$. Specializing the post-suppression ambient-OFDM model in \cite{Xu2024JointLocalization} to the single-BD case yields
\begin{equation}
	\widetilde{\mathbf H}_n
	=
	c_n\alpha
	\mathbf v\!\left(\theta(\mathbf p)\right)
	\mathbf d^T\!\left(\tau(\mathbf p)\right)
	+
	\mathbf W_n,
	\quad n=1,\ldots,N,
	\label{eq:cfr_observation}
\end{equation}
where
\begin{equation}
	\operatorname{vec}(\mathbf W_n)
	\sim
	\mathcal{CN}\!\left(
	\mathbf 0,
	\sigma^2\mathbf I_{MK_p}
	\right).
	\label{eq:effective_noise}
\end{equation}

Here, $\sigma^2$ is the per-entry noise variance in the adopted equivalent CFR model. Division by unit-modulus pilots preserves this variance. The ordinary transpose in \eqref{eq:cfr_observation} follows the steering convention: the $(m,\ell)$ entry of $\mathbf v(\theta)\mathbf d^T(\tau)$ is $v_m(\theta)e^{-j2\pi k_\ell\Delta f\tau}$.

Within a frame, $\mathbf p$, its induced delay--AoA geometry, and $\alpha$ remain fixed, whereas $c_n$ and $\mathbf W_n$ vary with $n$. We define the pre-combining input SNR per receive-antenna/pilot CFR coefficient as
\begin{equation}
	\mathrm{SNR}_{\mathrm{in}}
	=
	\frac{|\alpha|^2}{\sigma^2}.
	\label{eq:input_snr}
\end{equation}

\subsection{Joint Inference Problem}

Stacking the $N$ CFR matrices gives
\begin{equation}
	\boldsymbol{\mathcal H}
	=
	\operatorname{stack}_{n=1}^{N}
	\left\{
	\widetilde{\mathbf H}_n
	\right\}
	\in
	\mathbb C^{N\times M\times K_p}.
	\label{eq:stacked_cfr}
\end{equation}
The joint receiver is defined by the mapping
\begin{equation}
	\mathcal R:
	\mathbb C^{N\times M\times K_p}
	\longrightarrow
	\mathbb R^2\times\{0,1\}^{N-1},
	~
	(\widehat{\mathbf p},\widehat{\mathbf b})
	=
	\mathcal R(\boldsymbol{\mathcal H}),
	\label{eq:joint_receiver_mapping}
\end{equation}
where $\widehat{\mathbf p}\in\mathbb R^2$ is the continuous two-dimensional position estimate and
$\widehat{\mathbf b}=[\widehat b_1,\ldots,\widehat b_{N-1}]^T$ contains the detected information bits. Delay and AoA are position-induced intermediate variables rather than independent receiver outputs.

The source and receiver positions, array orientation, OFDM parameters, pilot locations, and pilot symbols are known, whereas $\mathbf p$, $\alpha$, and $\mathbf b$ are unknown. We assume a single dominant backscatter path, frame-static geometry and $\alpha$, OFDM synchronization, and a sufficiently long cyclic prefix.

In \eqref{eq:cfr_observation}, the geometry is shared across all $N$ symbols, whereas each information bit is associated with an adjacent-symbol transition. The receiver must therefore exploit repeated geometric evidence without coherently canceling responses modulated by opposite reflection signs, and retain competing hypotheses when the frame does not reliably distinguish them. Section~\ref{sec:proposed_receiver} develops a shared-posterior receiver for these coupled tasks.

\section{GeoFrameNet: Uncertainty-Preserving Joint Receiver}
\label{sec:proposed_receiver}

\subsection{Receiver Overview and Design Rationale}

\begin{figure*}[t]
	\centering
	\includegraphics[width=0.9\textwidth]{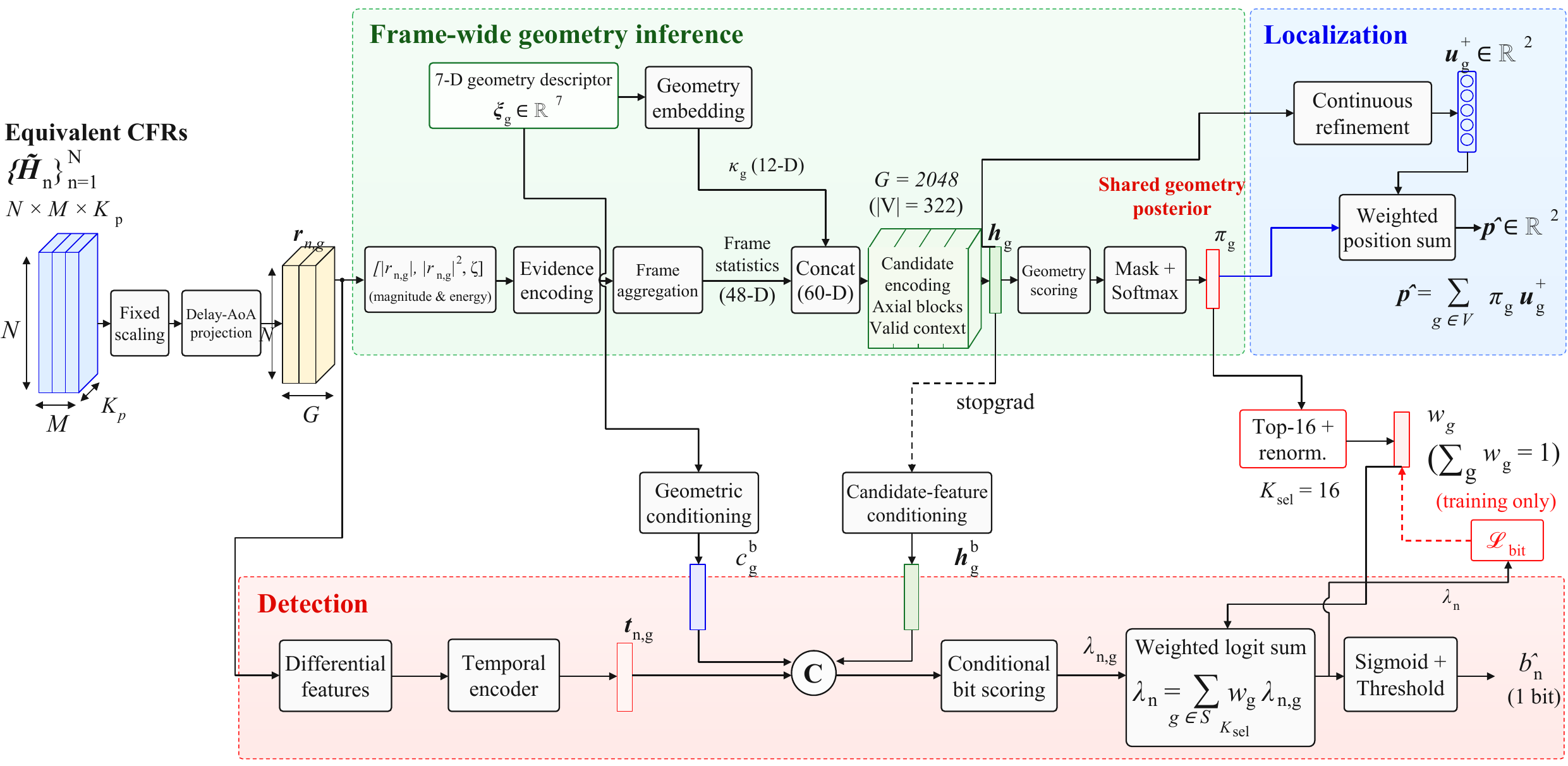}
	\caption{GeoFrameNet architecture. Frame-wide geometry inference produces a shared posterior for continuous localization and posterior-weighted differential detection. The circled C denotes feature concatenation.}
	\label{fig:geoframenet}
\end{figure*}

The central design choice is to strengthen geometry evidence without prematurely fixing the geometry used by the two tasks. Unlike the fixed-geometry detection stage in \cite{Xu2024JointLocalization}, GeoFrameNet uses a shared posterior formed from full-frame observations. Aggregation exploits repeated geometry evidence, while multi-hypothesis inference retains alternatives when that evidence remains ambiguous. As shown in Fig.~\ref{fig:geoframenet}, localization uses the full posterior with continuous refinement, whereas detection uses posterior weights renormalized over a top-$K_{\rm sel}$ subset. Communication supervision additionally guides geometry scoring through the posterior weights rather than the detached detector-content path.

\subsection{Shared-Geometry Posterior Construction}
\label{sec:shared_geometry_posterior}

A fixed affine scaling is applied component-wise to the equivalent CFR observation:
\begin{equation}
	\begin{aligned}
		\mathbf X_n
		={}
		\frac{
		\Re\{\widetilde{\mathbf H}_n\}-\mu_{\Re}
		}{
		\max(\sigma_{\Re},\epsilon_{\rm sc})
		}		
		+j
		\frac{
		\Im\{\widetilde{\mathbf H}_n\}-\mu_{\Im}
		}{
		\max(\sigma_{\Im},\epsilon_{\rm sc})
		}
		\in\mathbb C^{M\times K_p}.
	\end{aligned}
	\label{eq:method_scaled_cfr}
\end{equation}
Here, $\mu_{\Re}$ and $\mu_{\Im}$ are fixed scalar means, $\sigma_{\Re}$ and $\sigma_{\Im}$ are the corresponding standard deviations, and $\epsilon_{\rm sc}=10^{-15}$ prevents division by zero. The scaled matrices form an $N\times M\times K_p$ complex tensor, equivalently represented by separate real and imaginary channels of size $N\times M\times K_p\times2$.

Let
\[
\mathcal G
=
\left\{
(i,j):
i=1,\ldots,G_\theta,\;
j=1,\ldots,G_\tau
\right\}
\]
index the dense delay--AoA lattice.  The first computational axis is sampled uniformly in the spatial sine $s=\sin\theta$, not uniformly in $\theta$, and follows the wrapped spatial-frequency ordering
\begin{equation*}
	s_i
	=
	\begin{cases}
		\displaystyle
		\frac{2(i-1)}{G_\theta},
		&
		1\leq i\leq G_\theta/2,
		\\[1.2ex]
		\displaystyle
		\frac{2(i-1-G_\theta)}{G_\theta},
		&
		G_\theta/2<i\leq G_\theta,
	\end{cases}
\end{equation*}
and the normalized-delay samples are
\begin{equation*}
	\widetilde\tau_j
	=
	\frac{j-1}{G_\tau},
	\quad
	j=1,\ldots,G_\tau.
\end{equation*}
Here $G_\theta$ counts spatial-sine samples. We use $G_\theta=64$ and $G_\tau=32$, yielding $G=G_\theta G_\tau=2048$ lattice sites. For
$g=(i,j)\in\mathcal G$, define
\begin{equation*}
	s_g=s_i,
	\quad
	\widetilde\tau_g=\widetilde\tau_j,
	\quad
	\theta_i=\arcsin(s_i),
	\quad
	\tau_j
	=
	\frac{\widetilde\tau_j}{D_p\Delta f},
\end{equation*}
and the corresponding candidate atom
\begin{equation}
	\mathbf A_g
	=
	\mathbf v(\theta_i)
	\mathbf d^T(\tau_j).
	\label{eq:method_candidate_atom}
\end{equation}
All $G$ lattice sites are retained for candidate projection and local axial processing.

For each nondegenerate lattice site satisfying $L(\widetilde\tau_g)>d_{\mathsf{TR}}$, define the physical candidate coordinate and link distances as
\begin{equation*}
	\begin{aligned}
		\mathbf u_g
		&=
		\boldsymbol\psi(\widetilde\tau_g,s_g),
		\\
		d_{\mathsf{TB},g}
		&=
		\|\mathbf u_g-\mathbf p_{\mathsf T}\|_2,
		\quad
		d_{\mathsf{BR},g}
		&=
		\|\mathbf u_g-\mathbf p_{\mathsf R}\|_2.
	\end{aligned}
\end{equation*}
The nondegenerate physical posterior support is
\begin{equation*}
	\begin{aligned}
		\mathcal V
		=
		\{g\in\mathcal G:\;&
		L(\widetilde\tau_g)
		>
		d_{\mathsf{TR}}+\epsilon_L,\;
		\mathbf u_g\in\mathcal P,
		\\[-0.2ex]
		&
		d_{\mathsf{TB},g}\geq d_{\min},\;
		d_{\mathsf{BR},g}\geq d_{\min}
		\},
	\end{aligned}
\end{equation*}
where $d_{\min}=0.5$~m and $\epsilon_L=10^{-9}$~m. For the considered geometry, $|\mathcal V|=322$. This set defines the physical hypothesis support, while all $G=2048$ lattice sites are retained for dense feature processing.

The complex evidence for symbol $n$ and candidate $g$ is
\begin{equation}
	r_{n,g}
	=
	\frac{
		\langle\mathbf A_g,\mathbf X_n\rangle_{\mathrm F}
	}{
		\sqrt{MK_p}
	}
	=
	\frac{
		\mathbf v^H(\theta_i)
		\mathbf X_n
		\mathbf d^*(\tau_j)
	}{
		\sqrt{MK_p}
	},
	\label{eq:method_candidate_projection}
\end{equation}
where
\[
\langle\mathbf A,\mathbf H\rangle_{\mathrm F}
=
\operatorname{tr}(\mathbf A^H\mathbf H).
\]
Thus, $r_{n,g}$ quantifies the complex spatial--frequency evidence associated with candidate $g$ in symbol $n$ and is retained for differential detection.

Frame-shared geometry does not imply a frame-shared complex response: the backscatter component is modulated by the DBPSK symbol $c_n$. Direct coherent averaging can therefore cancel useful signal components. GeoFrameNet separates the two uses of the observations: magnitude- and energy-based evidence is aggregated over the frame for geometry inference, while the complex projections $r_{n,g}$ are retained for differential detection. This separation exploits repeated geometry observations without first deciding the unknown reflection signs or discarding the phase information needed for bit recovery.

Define the fixed affine position normalization
\begin{equation}
	\mathcal N_p(\mathbf x)
	\triangleq
	\mathbf L_p^{-1}
	(\mathbf x-\boldsymbol\mu_p).
	\label{eq:method_position_normalization}
\end{equation}
Here, $\boldsymbol\mu_p$ is a fixed coordinate center and $\mathbf L_p$ is an invertible coordinate-scaling matrix. The reference distance $d_{\rm ref}$ normalizes the link-distance features. Their numerical values are specified in Section~\ref{sec:simulation_results}.

To retain the regular $G_\theta\times G_\tau$ tensor, every lattice site is assigned a finite deterministic auxiliary coordinate. Define
\begin{equation*}
	\begin{aligned}
		D_g
		&=
		2\!\left[
		L(\widetilde\tau_g)
		+\boldsymbol\Delta_{\mathsf{TR}}^T\mathbf u(s_g)
		\right],\\
		D_{g,\epsilon}
		&=
		\operatorname{sgn}_{+}(D_g)
		\max\{|D_g|,\epsilon_D\},
		\quad
		\epsilon_D=10^{-9}~{\rm m},\\
		r_g^{\rm lat}
		&=
		\operatorname{clip}\!\left(
		\frac{
			L^2(\widetilde\tau_g)
			-\|\boldsymbol\Delta_{\mathsf{TR}}\|_2^2
		}{
			D_{g,\epsilon}
		},
		0,4d_{\mathsf{TR}}
		\right),
	\end{aligned}
\end{equation*}
where $\operatorname{sgn}_{+}(x)=1$ for $x\geq0$ and $\operatorname{sgn}_{+}(x)=-1$ otherwise, and
$\operatorname{clip}(x,a,b)=\min\{\max\{x,a\},b\}$. The auxiliary coordinate is
$\mathbf u_g^{\rm lat}
=\mathbf p_{\mathsf R}+r_g^{\rm lat}\mathbf u(s_g)$, with
$d_{\mathsf{TB},g}^{\rm lat}
=\|\mathbf u_g^{\rm lat}-\mathbf p_{\mathsf T}\|_2$
and
$d_{\mathsf{BR},g}^{\rm lat}
=\|\mathbf u_g^{\rm lat}-\mathbf p_{\mathsf R}\|_2$.
No Cartesian clipping to $\mathcal P$ is applied. For valid candidates in the considered configuration, the guard and radial clip are inactive, so the auxiliary and physical quantities coincide. Invalid-site descriptors remain available to intermediate lattice processing but are excluded from valid-support context aggregation, posterior normalization and supervision, localization summation, and top-$K_{\rm sel}$ selection.

Each lattice site is associated with the deterministic seven-dimensional descriptor
\begin{equation}
	\boldsymbol\xi_g
	=
	\left[
	s_g,
	\sqrt{1-s_g^2},
	2\widetilde\tau_g-1,
	\mathcal N_p(\mathbf u_g^{\rm lat})^T,
	\frac{d_{\mathsf{TB},g}^{\rm lat}}{d_{\rm ref}},
	\frac{d_{\mathsf{BR},g}^{\rm lat}}{d_{\rm ref}}
	\right]^T.
	\label{eq:method_coordinate_features}
\end{equation}
The descriptor is determined by the candidate index and known system geometry, not by the unknown BD position. For $g\in\mathcal V$, its coordinate and distance components correspond to a physically feasible candidate.

The frame-scale descriptor is

\begin{equation}
	\zeta
	=
	\log\!\left(
	\sqrt{
		\max\!\left[
		\frac{1}{NG}
		\sum_{n=1}^{N}
		\sum_{g\in\mathcal G}
		|r_{n,g}|^2,
		\epsilon_\zeta
		\right]
	}
	\right).
	\label{eq:method_log_rms}
\end{equation}
Here $\epsilon_\zeta=10^{-12}$ is the floor applied to the mean squared evidence before the square root and logarithm. For each symbol and lattice site,
\begin{equation}
	\mathbf e_{n,g}
	=
	f_{\rm e}
	\left(
	[|r_{n,g}|,|r_{n,g}|^2,\zeta]^T
	\right)
	\in\mathbb R^{12},
	\label{eq:method_symbol_evidence}
\end{equation}
where $f_{\rm e}$ is a multilayer perceptron (MLP) with two 32-neuron hidden layers and Gaussian error linear unit (GELU) activations. In parallel, the seven-dimensional descriptor $\boldsymbol\xi_g$ is mapped to $\boldsymbol\kappa_g\in\mathbb R^{12}$ by an MLP with one 48-neuron hidden layer and GELU activation, followed by layer normalization.

The symbol-wise evidence is aggregated for each lattice site as
\begin{equation}
	\begin{aligned}
		\boldsymbol\mu_g
		&=
		\frac{1}{N}
		\sum_{n=1}^{N}
		\mathbf e_{n,g},
		&
		\boldsymbol\nu_g
		&=
		\frac{1}{N}
		\sum_{n=1}^{N}
		\mathbf e_{n,g}^{\odot2},
		\\
		\mathbf m_g
		&=
		\max_{1\leq n\leq N}
		\mathbf e_{n,g},
		\\
		\boldsymbol\sigma_g^{(e)}
		&=
		\sqrt{
			\frac{1}{N}
			\sum_{n=1}^{N}
			(\mathbf e_{n,g}-\boldsymbol\mu_g)^{\odot2}
			+10^{-8}
		}.
	\end{aligned}
	\label{eq:method_frame_statistics}
\end{equation}
Here, $(\cdot)^{\odot2}$ denotes elementwise squaring, and the maximum is taken elementwise over the $N$ symbols. These statistics summarize the empirical mean, second moment, maximum, and variability of each learned evidence channel across the frame, without coherently summing the complex responses.

Concatenating these four 12-dimensional statistics with $\boldsymbol\kappa_g$ yields a 60-dimensional candidate summary. An MLP with a 128-neuron hidden layer maps this summary to a 64-dimensional feature, followed by layer normalization. The resulting 64-channel features are arranged on the $G_\theta\times G_\tau$ lattice using the prescribed spatial-sine and normalized-delay ordering. Axial processing exchanges information along these indexed dimensions before global context is aggregated over the valid support.

Four residual axial blocks apply depthwise $7\times1$ and $1\times5$ convolutions along the spatial-frequency and delay axes, respectively. Their outputs are summed and mixed by a $1\times1$ convolution with a residual connection and group normalization, followed by a second residual channel-mixing block with a 256-channel hidden expansion.

Let $\widetilde{\mathbf h}_g\in\mathbb R^{64}$ denote the resulting local lattice feature. While the axial blocks capture local candidate-neighborhood structure, candidate scoring can also depend on the global support pattern. GeoFrameNet therefore incorporates context over the physical posterior support through
\begin{equation}
	\begin{aligned}
		\overline{\mathbf h}
		&=
		\frac{1}{|\mathcal V|}
		\sum_{g\in\mathcal V}
		\widetilde{\mathbf h}_g,
		\\
		\mathbf a
		&=
		f_{\rm glb}(\overline{\mathbf h}),
		\quad
		\mathbf h_g
		=
		\widetilde{\mathbf h}_g
		+\mathbf a
		+f_{\rm cap}(\mathbf a),
	\end{aligned}
	\label{eq:method_candidate_latent}
\end{equation}
where $f_{\rm glb}$ and $f_{\rm cap}$ each use two 256-neuron hidden layers and produce 64-dimensional outputs; layer normalization is applied at the output of $f_{\rm glb}$. The resulting candidate feature
$\mathbf h_g\in\mathbb R^{64}$ combines local lattice evidence with global valid-candidate context and is shared by the posterior, refinement, and detection paths.

A GELU-activated MLP with a 64-neuron hidden layer maps $\mathbf h_g$ to the scalar geometry logit $\ell_g$. After masking sites outside $\mathcal V$, a softmax gives
\begin{equation}
	\pi_g
	=
	\frac{\exp(\ell_g)}
	{\sum_{j\in\mathcal V}\exp(\ell_j)},
	\quad g\in\mathcal V,
	\quad
	\pi_g=0,
	\quad g\notin\mathcal V.
	\label{eq:method_geometry_posterior}
\end{equation}
The vector $\boldsymbol\pi=[\pi_g]_{g\in\mathcal G}$ is the learned geometry posterior, normalized over $\mathcal V$ and zero elsewhere. It retains candidate-level weights for the two receiver tasks without imposing an intermediate top-1 geometry decision.

\subsection{Task-Specific Posterior Inference and Controlled Coupling}

Full-frame aggregation does not guarantee that a single geometry candidate is reliably identified. Conditioning all bit decisions on the highest-scoring candidate would reintroduce a premature geometry commitment. GeoFrameNet therefore retains the full posterior for localization and combines candidate-conditioned differential logits over a top-$K_{\rm sel}$ subset. Truncation limits candidate-specific temporal processing, while renormalized posterior weights allow multiple plausible geometries to contribute to each bit decision. Let $\mathcal S_{K_{\rm sel}}\subset\mathcal V$ denote the indices of the $K_{\rm sel}$ largest posterior masses among the valid candidates. The selected masses are renormalized as
\begin{equation}
	w_g
	=
	\frac{\pi_g}
	{\sum_{j\in\mathcal S_{K_{\rm sel}}}\pi_j},
	\quad
	g\in\mathcal S_{K_{\rm sel}}.
	\label{eq:method_topk_weights}
\end{equation}
Candidates outside $\mathcal S_{K_{\rm sel}}$ remain in the full posterior used for localization. We set $K_{\rm sel}=16$ in all experiments.

Detection is motivated by the differential structure in \eqref{eq:dbpsk_mapping}. Before the fixed affine preprocessing in \eqref{eq:method_scaled_cfr}, the noiseless matched response of a well-matched candidate is proportional to $c_n\alpha$, and hence
\begin{equation*}
	(c_{n+1}\alpha)
	(c_n\alpha)^*
	=
	c_{n+1}c_n|\alpha|^2
	=
	(1-2b_n)|\alpha|^2.
\end{equation*}
Thus, the frame-static composite phase cancels, while the differential sign carries the information bit.

For the transition from symbol $n$ to symbol $n+1$ and candidate $g$, define
\begin{equation*}
	z_{n,g}
	=
	r_{n+1,g}r_{n,g}^*,
	\quad
	a_{n,g}
	=
	|r_{n+1,g}||r_{n,g}|.
\end{equation*}
The six-dimensional transition feature is
\begin{equation}
	\begin{aligned}
		\boldsymbol\chi_{n,g}
		=
		\big[
		&
		\Re\{z_{n,g}\}/\bar a_{n,g},
		\Im\{z_{n,g}\}/\bar a_{n,g},
		\\
		&
		\log(1+|z_{n,g}|),
		\log(1+|r_{n+1,g}|),
		\\
		&
		\log(1+|r_{n,g}|),
		\log(1+a_{n,g})
		\big]^T,
	\end{aligned}
	\label{eq:method_transition_features}
\end{equation}
where $\bar a_{n,g}=\max(a_{n,g},10^{-8})$. The normalized complex product captures the adjacent-symbol differential phase/sign relation, while the logarithmic terms provide magnitude context. Because $|z_{n,g}|=a_{n,g}$, the third and sixth entries are algebraically identical and are both retained in the six-dimensional detector input. The first transition uses the fixed reference $c_1=+1$ through $r_{1,g}$.

A GELU-activated MLP with a 64-neuron hidden layer maps the six-dimensional transition feature to a 32-dimensional feature. Two residual temporal blocks with kernel size five then process the transition sequence to produce $\mathbf t_{n,g}\in\mathbb R^{32}$. Each temporal block comprises a 32-channel one-dimensional convolution, GELU activation, a pointwise convolution, a residual connection, and layer normalization. The detector additionally uses a 16-dimensional geometry feature $\mathbf c_g^{\rm b}$ obtained by applying an MLP with a 32-neuron hidden layer to $\boldsymbol\xi_g$, and a 32-dimensional candidate feature $\mathbf h_g^{\rm b}$ obtained by applying an MLP with a 64-neuron hidden layer to $\operatorname{stopgrad}(\mathbf h_g)$. Here, $\operatorname{stopgrad}(\cdot)$ is the identity in the forward pass and blocks gradients during backpropagation. The circled C in Fig.~\ref{fig:geoframenet} concatenates the three feature vectors into an 80-dimensional input ($32+16+32$), which is processed as
\begin{equation} 
	\lambda_{n,g}=f_{\rm b}\!\left(\left[\mathbf t_{n,g}^{T},(\mathbf c_g^{\rm b})^{T},(\mathbf h_g^{\rm b})^{T}\right]^{T}\right),
	\label{eq:method_conditional_logit} 
	\end{equation}
where $f_{\rm b}$ is an MLP with hidden layers of 64 and 32 neurons and a scalar output, yielding the candidate-conditioned bit logit $\lambda_{n,g}$.

%The gradient structure provides controlled coupling between localization and detection. Blocking the gradient path through the posterior weights would prevent the bit-detection loss from influencing geometry scoring, whereas allowing the loss to propagate from the detector back into the shared candidate feature $\mathbf h_g$ would directly modify the common geometry representation. GeoFrameNet therefore applies $\operatorname{stopgrad}(\cdot)$ to $\mathbf h_g$ before the candidate-feature MLP, while retaining the gradient path through the posterior-derived weights $w_g$. Consequently, the bit-detection loss can update the detector parameters and the geometry-posterior network, but cannot update the shared candidate feature $\mathbf h_g$ through this detector branch.

The final bit logit and probability are
\begin{equation}
	\begin{aligned}
		\lambda_n
		&=
		\sum_{g\in\mathcal S_{K_{\rm sel}}}
		w_g\lambda_{n,g},
		\\
		\rho_n
		&=
		\operatorname{sigmoid}(\lambda_n),
		\quad
		n=1,\ldots,N-1.
	\end{aligned}
	\label{eq:method_bit_mixture}
\end{equation}
Let $\boldsymbol\lambda=[\lambda_1,\ldots,\lambda_{N-1}]^T$. The hard decision is $\widehat b_n=\mathbf{1}\{\rho_n\geq0.5\}$, where $\mathbf{1}\{\cdot\}$ denotes the indicator function.

Geometry-only supervision does not directly optimize the weighting of candidate-conditioned logits for bit detection. We therefore let the bit loss act on the final logits $\boldsymbol\lambda$ and influence geometry scoring through the posterior weights in \eqref{eq:method_bit_mixture}. The operation $\operatorname{stopgrad}(\mathbf h_g)$ separates this weighting feedback from direct adaptation of the shared representation through the detector-content path; detector parameters still receive bit-loss gradients. The shared representation can be updated through the weights, the selected top-$K_{\rm sel}$ indices are treated as fixed during backpropagation, and the bit loss does not update the refinement-MLP parameters.

\subsection{Continuous Refinement and Training Objective}
\label{sec:geoframe_refinement}

Posterior weighting can combine several hypotheses, but it does not adapt their fixed reference coordinates to the observation. Although a weighted sum of lattice coordinates can already be continuous, candidate-specific refinement reduces local lattice mismatch without increasing the candidate count. GeoFrameNet therefore predicts bounded delay and spatial-sine corrections before position averaging. Without refinement, the posterior yields the coarse estimate
\[
\widehat{\mathbf p}_{\rm coarse}
=
\sum_{g\in\mathcal V}
\pi_g\mathbf u_g.
\]

The refinement MLP maps the shared candidate feature $\mathbf h_g$ to two scalars, $o_{\widetilde\tau,g}$ and $o_{s,g}$, using a 64-neuron hidden layer with GELU activation. These outputs parameterize bounded corrections in normalized delay and spatial sine:
\begin{equation}
	\delta_{\widetilde\tau,g}
	=
	\frac{1}{64}\tanh( o_{\widetilde\tau,g}),
	\quad
	\delta_{s,g}
	=
	\frac{1}{64}\tanh(o_{s,g}).
	\label{eq:method_bounded_offsets}
\end{equation}
Both learned offsets are bounded by half the nominal lattice spacing before the feasibility clipping below.

Define the normalized nondegenerate delay boundary as
\[
\widetilde\tau_{\min}
=
D_p\Delta f\,
\frac{d_{\mathsf{TR}}}{c_0},
\]
and let $\epsilon_\tau=10^{-6}$ denote a numerical margin in the normalized-delay domain. The refined geometric parameters are
\begin{equation}
	\begin{aligned}
		\widetilde\tau_g^+
		&=
		\operatorname{clip}
		\left(
		\widetilde\tau_g+\delta_{\widetilde\tau,g},
		\widetilde\tau_{\min}+\epsilon_\tau,
		1-10^{-7}
		\right),
		\\
		s_g^+
		&=
		\operatorname{clip}
		\left(
		s_g+\delta_{s,g},
		-1+10^{-5},
		1-10^{-5}
		\right).
	\end{aligned}
	\label{eq:method_refined_geometry}
\end{equation}

The delay clip excludes the degenerate bistatic boundary, while the spatial-sine clip keeps $s_g^+$ strictly inside $(-1,1)$. These parameter constraints do not project the refined Cartesian coordinates onto $\mathcal P$.

Applying the visible-branch inverse in \eqref{eq:visible_branch_inverse} gives the refined candidate
coordinate and final position estimate:
\begin{equation}
	\mathbf u_g^+
	=
	\boldsymbol\psi(
	\widetilde\tau_g^+,
	s_g^+),
	\quad
	\widehat{\mathbf p}
	=
	\sum_{g\in\mathcal V}
	\pi_g\mathbf u_g^+.
	\label{eq:method_position_readout}
\end{equation}
Candidate-specific refinement addresses local lattice mismatch, while the full posterior is retained until the weighted position estimate is formed. No additional projection onto $\mathcal P$ is applied to the final estimate.

Training jointly optimizes bit and position recovery together with geometry-aware auxiliary objectives, posterior supervision, and candidate-specific refinement. Let $\ell_{\rm H}(\cdot,\cdot)$ denote the elementwise smooth-$L_1$ loss with unit transition parameter $\beta_{\rm H}=1$, and let $\ell_{\rm BCE}(\cdot,\cdot)$ denote binary cross-entropy evaluated from logits. The bit loss is
\begin{equation}
	\mathcal L_{\rm bit}
	=
	\ell_{\rm BCE}
	(\boldsymbol\lambda,\mathbf b).
	\label{eq:method_bit_loss}
\end{equation}
Here, $\ell_{\rm BCE}$ is averaged over the mini-batch and the $N-1$ information-bit transitions. In
\eqref{eq:method_geometry_losses}, $\operatorname{mean}$ denotes averaging over the mini-batch and, for vector-valued quantities, over their components.

The direct position and geometry-aware auxiliary losses are
\begin{equation}
	\begin{aligned}
		\mathcal L_{\rm pos}
		&=
		\operatorname{mean}\,
		\ell_{\rm H}
		\left(
		\mathcal N_p(\widehat{\mathbf p}),
		\mathcal N_p(\mathbf p)
		\right),
		\\
		\mathcal L_{\tau}
		&=
		\operatorname{mean}\,
		\ell_{\rm H}
		\left(
		\widehat{\widetilde\tau},
		\widetilde\tau
		\right),
		\\
		\mathcal L_s
		&=
		\operatorname{mean}\,
		\ell_{\rm H}
		\left(
		\sin\theta(\widehat{\mathbf p}),
		\sin\theta(\mathbf p)
		\right),
	\end{aligned}
	\label{eq:method_geometry_losses}
\end{equation}
where
\[
\widehat{\widetilde\tau}
=
D_p\Delta f\,\tau(\widehat{\mathbf p}),
\quad
\widetilde\tau
=
D_p\Delta f\,\tau(\mathbf p).
\]
The delay and spatial-sine terms are obtained from the predicted and true positions through the same deterministic geometry mapping and therefore serve as auxiliary constraints rather than independent receiver outputs.

For posterior supervision, a smooth target is defined on the valid physical candidates using their unrefined coordinates:
\begin{equation} 
	\begin{aligned} \widetilde q_g(\mathbf p)&=\exp\!\left(-\frac{\|\mathbf p-\mathbf u_g\|_2^2}{2\sigma_a^2}\right),\quad g\in\mathcal V,\\ q_g(\mathbf p)&=\frac{\widetilde q_g(\mathbf p)}{\sum_{j\in\mathcal V}\widetilde q_j(\mathbf p)},\quad g\in\mathcal V,\quad \sigma_a=0.40~\mathrm{m}. 
	\end{aligned} 
	\label{eq:method_soft_anchor_target} 
\end{equation}

For $g\notin\mathcal V$, set $\widetilde q_g(\mathbf p)=q_g(\mathbf p)=0$. This position-dependent distribution is a training target and is not required during inference.

Define the valid-support vectors
\[
\mathbf q_{\mathcal V}(\mathbf p)
=
[q_g(\mathbf p)]_{g\in\mathcal V},
\quad
\boldsymbol\pi_{\mathcal V}
=
[\pi_g]_{g\in\mathcal V}.
\]
The posterior and refinement losses are
\begin{equation}
	\begin{aligned}
		\mathcal L_{\rm post}
		&=
		\operatorname{avg}_{\mathcal B}
		D_{\rm KL}
		\!\left(
		\mathbf q_{\mathcal V}(\mathbf p)
		\,\|\,\boldsymbol\pi_{\mathcal V}
		\right),
		\\
		\mathcal L_{\rm off}
		&=
		\operatorname{avg}_{\mathcal B}
		\sum_{g\in\mathcal V}
		q_g(\mathbf p)
		\operatorname{avg}_{x,y}
		\ell_{\rm H}
		\left(
		\mathbf u_g^+/d_{\rm ref},
		\mathbf p/d_{\rm ref}
		\right).
	\end{aligned}
	\label{eq:method_posterior_offset_losses}
\end{equation}
Here, $\mathcal B$ denotes the current mini-batch, $\operatorname{avg}_{\mathcal B}$ its empirical average, $\operatorname{avg}_{x,y}$ averaging over the two Cartesian coordinates, and $D_{\rm KL}$ the Kullback--Leibler (KL) divergence. The posterior loss supervises candidate scores using geometric proximity, whereas the offset loss supervises individual refined coordinates with target weights $q_g(\mathbf p)$ rather than predicted weights $\pi_g$. It therefore complements the loss on the final posterior-weighted position estimate.

The total training objective is
\[
	\mathcal L
	=
	\mathcal L_{\rm bit}
	+2\mathcal L_{\rm pos}
	+0.25\mathcal L_{\tau}
	+0.25\mathcal L_s
	+0.5\mathcal L_{\rm post}
	+0.10\mathcal L_{\rm off}.
\]

At inference, GeoFrameNet produces $\widehat{\mathbf p}$ and $\widehat{\mathbf b}$ in one fixed-depth forward pass without an intermediate geometry point estimate.

\begin{figure*}[t]
	\centering
	\includegraphics[width=0.8\textwidth]{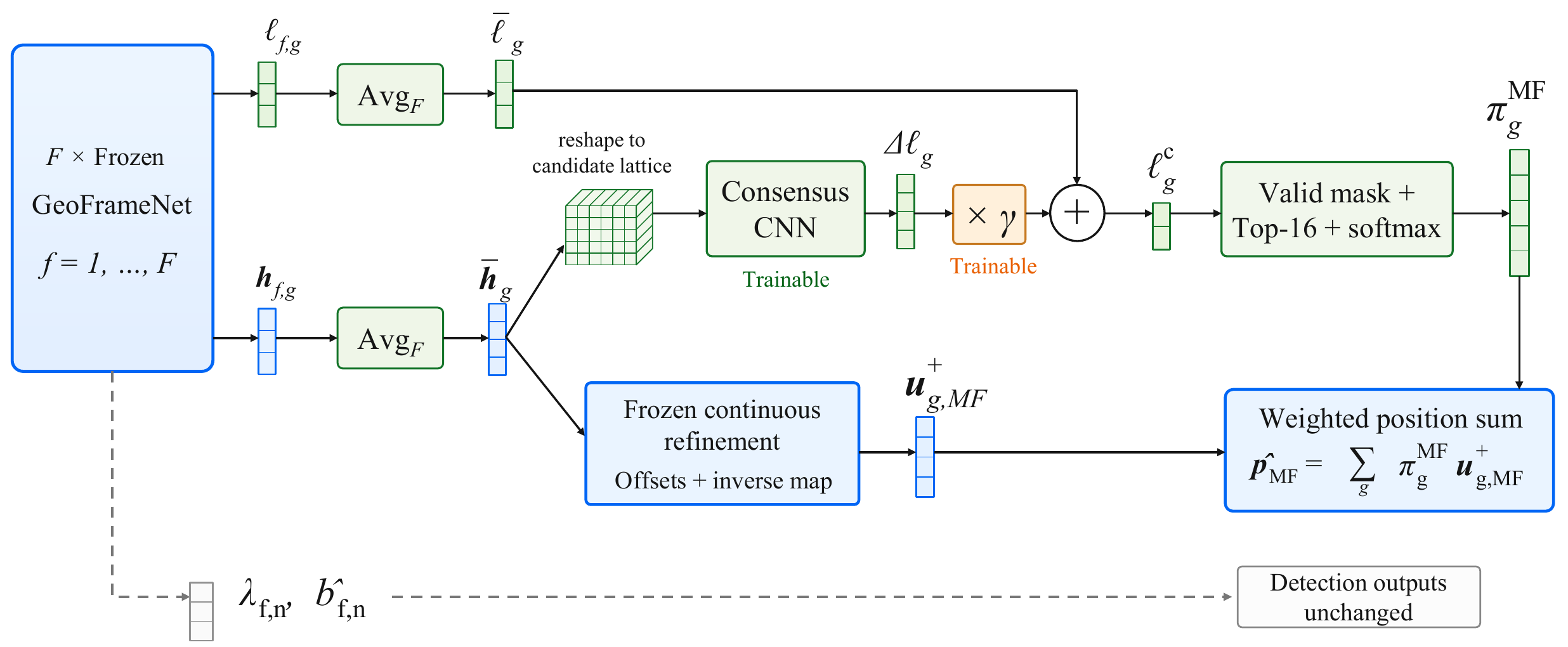}
	\caption{CFGC-Net architecture. Candidate-level fusion enhances
		cross-frame localization while preserving all frame-wise
		detection outputs.}
	\label{fig:cfgc}
\end{figure*}

\section{CFGC-Net: Cross-Frame Geometry Consensus for Localization Enhancement}
\label{sec:multiframe_extension}

\subsection{Motivation and Design Overview}
\label{sec:mf_overview}

We consider $F$ frames from a BD at a fixed position. Geometry and path-loss magnitude are shared, whereas payloads, composite phases, and noise realizations are frame specific. Direct complex-CFR averaging can therefore cause destructive combination, while averaging final Cartesian estimates discards candidate-level information before fusion.

As shown in Fig.~\ref{fig:cfgc}, CFGC-Net combines the geometry logits and candidate features produced by frozen GeoFrameNet passes. Mean logits provide a parameter-free reference; averaged features support both consensus-score correction and frozen continuous refinement. Cross-frame fusion is restricted to localization, and all frame-wise detection outputs are retained unchanged.

\subsection{Candidate-Level Cross-Frame Consensus}
\label{sec:mf_consensus_adapter}

For each common lattice index $g\in\mathcal G$, the frozen single-frame receiver supplies a pre-softmax geometry logit $\ell_{f,g}$ and candidate feature $\mathbf h_{f,g}\in\mathbb R^{64}$. CFGC-Net forms a permutation-invariant summary by averaging corresponding logits and features across the $F$ frames:
\begin{equation}
	\overline\ell_g
	=
	\frac{1}{F}
	\sum_{f=1}^{F}
	\ell_{f,g},
	\quad
	\overline{\mathbf h}_g
	=
	\frac{1}{F}
	\sum_{f=1}^{F}
	\mathbf h_{f,g}.
	\label{eq:mf_equal_means}
\end{equation}
The mean geometry logit $\overline\ell_g$ serves as a parameter-free reference for cross-frame consensus.

The averaged candidate feature vectors are arranged on a 64-channel $G_\theta\times G_\tau$ delay--AoA lattice $\overline{\boldsymbol{\mathcal Z}}\in\mathbb R^{64\times G_\theta\times G_\tau}$.  Here the superscript $(\theta)$ on the axial convolution denotes the wrapped spatial-sine-indexed axis, rather than uniform sampling in physical angle.
A lightweight axial refiner exchanges information along the two physical candidate dimensions:
\begin{equation}
	\boldsymbol{\mathcal Z}_{\Delta}
	=
	\mathcal C_{1\times1}
	\!\left(
	\operatorname{GELU}
	\!\left(
	\mathcal C_{3\times1}^{(\theta)}
	(\overline{\boldsymbol{\mathcal Z}})
	+
	\mathcal C_{1\times3}^{(\tau)}
	(\overline{\boldsymbol{\mathcal Z}})
	\right)
	\right),
	\label{eq:mf_axial_refiner}
\end{equation}
where the two axial convolutions preserve 64 channels and $\mathcal C_{1\times1}$ maps the fused features to one residual logit per candidate. Let $\Delta\ell_g$ denote the residual logit in $\boldsymbol{\mathcal Z}_{\Delta}$ associated with candidate $g$. With a learned scalar gate $\gamma$, the corrected consensus score is
\begin{equation}
	\ell_g^{\rm c}
	=
	\overline\ell_g
	+
	\gamma\Delta\ell_g.
	\label{eq:mf_consensus_logit}
\end{equation}
Thus, the trainable consensus module learns only a feature-based residual correction to the parameter-free mean-logit reference.

After masking invalid candidates, the $K_{\rm sel}=16$ largest valid corrected logits are retained to form the support $\mathcal S_{\rm MF}$. The corresponding consensus posterior is
\begin{equation}
	\pi_g^{\rm MF}
	=
	\begin{cases}
		\displaystyle
		\frac{\exp(\ell_g^{\rm c})}
		{\sum_{j\in\mathcal S_{\rm MF}}
			\exp(\ell_j^{\rm c})},
		& g\in\mathcal S_{\rm MF},
		\\[1ex]
		0,
		& g\notin\mathcal S_{\rm MF}.
	\end{cases}
	\label{eq:mf_sparse_posterior}
\end{equation}
Unlike single-frame localization, which uses the full posterior over $\mathcal V$, the cross-frame estimator uses the truncated posterior on $\mathcal S_{\rm MF}\subset\mathcal V$.

For posterior supervision, the smooth target $q_g(\mathbf p)$ in \eqref{eq:method_soft_anchor_target} is restricted to $\mathcal S_{\rm MF}$ and renormalized as
\begin{equation*}
	q_{g,\mathcal S}^{\rm MF}(\mathbf p)
	=
	\frac{q_g(\mathbf p)}
	{\sum_{j\in\mathcal S_{\rm MF}}q_j(\mathbf p)},
	\quad
	g\in\mathcal S_{\rm MF}.
\end{equation*}
Define the selected-support vectors
\[
\mathbf q_{\mathcal S}^{\rm MF}(\mathbf p)
=
[q_{g,\mathcal S}^{\rm MF}(\mathbf p)]_
{g\in\mathcal S_{\rm MF}},
\quad
\boldsymbol\pi_{\mathcal S}^{\rm MF}
=
[\pi_g^{\rm MF}]_{g\in\mathcal S_{\rm MF}}.
\]
The supervision target and predicted posterior are normalized on the same selected support. The selected indices are treated as fixed during backpropagation, while gradients pass through the softmax values of the selected corrected logits.

\subsection{Continuous Position Estimation and Training Objective}
\label{sec:mf_training_scope}

The averaged feature $\overline{\mathbf h}_g$ is passed through the frozen GeoFrameNet refinement MLP. The bounded offsets and clipping in \eqref{eq:method_bounded_offsets}--\eqref{eq:method_refined_geometry}, followed by the inverse mapping in \eqref{eq:visible_branch_inverse}, produce $\mathbf u_{g,\rm MF}^{+}$. Thus, CFGC-Net refines averaged candidate features rather than averaging independently refined Cartesian coordinates.

The resulting cross-frame position estimate is
\begin{equation}
	\widehat{\mathbf p}_{\rm MF}
	=
	\sum_{g\in\mathcal S_{\rm MF}}
	\pi_g^{\rm MF}
	\mathbf u_{g,\rm MF}^{+}.
	\label{eq:mf_position_readout}
\end{equation}
CFGC-Net uses the same bounded inverse mapping and unprojected Cartesian coordinate estimate as GeoFrameNet.

Only the axial consensus refiner and scalar gate in \eqref{eq:mf_axial_refiner}--\eqref{eq:mf_consensus_logit} are trainable. The effective training objective is
\begin{equation}
	\mathcal L_{\rm MF}
	=
	2\mathcal L_{\rm pos}^{\rm MF}
	+
	0.25\mathcal L_{\tau}^{\rm MF}
	+
	0.25\mathcal L_s^{\rm MF}
	+
	0.5\mathcal L_{\rm post}^{\rm MF},
	\label{eq:mf_training_loss}
\end{equation}
where the position, delay, and spatial-sine terms follow \eqref{eq:method_geometry_losses} with $\widehat{\mathbf p}_{\rm MF}$ as the predicted position. The posterior term, evaluated on $\mathcal S_{\rm MF}$, is
\begin{equation*}
	\mathcal L_{\rm post}^{\rm MF}
	=
	\operatorname{avg}_{\mathcal B}
	D_{\rm KL}
	\!\left(
	\mathbf q_{\mathcal S}^{\rm MF}(\mathbf p)
	\,\|\, 
	\boldsymbol\pi_{\mathcal S}^{\rm MF}
	\right).
\end{equation*}

CFGC-Net introduces $24{,}770$ trainable consensus parameters while reusing the frozen GeoFrameNet. Frame-specific communication outputs remain identical to those of the corresponding frozen single-frame passes, i.e.,
$\lambda_{f,n}^{\rm MF}=\lambda_{f,n}$,
with unchanged hard decisions. No bit voting or cross-frame data decoding is performed.

Equal frame averaging makes the fusion permutation invariant to frame ordering for each evaluated $F$. The method assumes a common BD geometry across the fused frames but does not require repeated communication payloads; generalization to frame counts outside the evaluated set is not assumed.

\section{Simulation Results and Analysis}
\label{sec:simulation_results}

\subsection{Simulation Setup}

The ambient source and receiver are located at $(0,0)$~m and $(20,0)$~m, respectively, with ULA-axis and positive-broadside directions $\mathbf u_a=[1,0]^T$ and $\mathbf u_b=[0,1]^T$. The BD position is uniformly sampled over $[0,20]^2~\mathrm{m}^2$, excluding locations within $0.5$~m of either endpoint. The source--BD and BD--receiver power gains are $L_{\mathsf{TB}}(\mathbf p)=10^{-3}d_{\mathsf{TB}}^{-2}(\mathbf p)$ and $L_{\mathsf{BR}}(\mathbf p)=10^{-3}d_{\mathsf{BR}}^{-2}(\mathbf p)$, respectively. The composite backscatter phase is independently drawn for each frame.

The simulator directly generates the equivalent post-suppression CFR in \eqref{eq:cfr_observation} with circular complex additive white Gaussian noise (AWGN); environmental acquisition and suppression are not simulated explicitly. Table~\ref{tab:simulation_settings} summarizes the system parameters. Position normalization uses $\boldsymbol\mu_p=[10,10]^T$~m, $\mathbf L_p=10\mathbf I_2$~m, and $d_{\rm ref}=20$~m. The frozen CFR scaler has the rounded values $(\mu_{\Re},\mu_{\Im}) \approx(-1.908\times10^{-9},-2.101\times10^{-8})$ and $(\sigma_{\Re},\sigma_{\Im}) \approx(1.494,1.493)\times10^{-4}$.

\begin{table}[t]
	\caption{Key Simulation Parameters}
	\label{tab:simulation_settings}
	\centering
	\footnotesize
	\setlength{\tabcolsep}{3pt}
	\begin{tabular}{p{0.38\columnwidth}p{0.51\columnwidth}}
		\hline
		Parameter & Value \\
		\hline
		Receive array
		& $M=32$, $d_a/\lambda_c=0.5$ \\
		
		OFDM
		& $K=64$, $\Delta f=1$~MHz, 64~MHz bandwidth \\
		
		Pilots
		& $K_p=16$, $D_p=4$ \\
		
		Frame
		& $N=64$ symbols, 63 information bits \\
		
		Candidate lattice / posterior support & $64\times32=2048$ sites, $|\mathcal V|=322$, $K_{\rm sel}=16$ \\
		\hline
	\end{tabular}
\end{table}

Training, validation, and test data use mutually disjoint random seeds, and test data are excluded from model and threshold selection. GeoFrameNet is trained using AdamW with a learning rate of $2\times10^{-4}$, weight decay $10^{-5}$, batch size 8, and gradient clipping at 5. CFGC-Net freezes GeoFrameNet and trains its 24,770 consensus parameters over $F\in\{2,4,8\}$ using the same optimizer settings.

For single-frame evaluation, the same 300 underlying test frames are reused across the primary SNR points. The BD position, payload, composite coefficient, noiseless CFR, and normalized noise realization of each frame remain fixed across SNR; only the AWGN variance is rescaled. Multi-frame
evaluation uses a separate set of 300 groups per SNR. Within each group, the BD position and path-loss magnitude are shared, whereas payloads, composite phases, and noise realizations are independently generated across frames. The $F=1,2,4,8$ cases use nested prefixes of the same eight-frame group.

\subsection{Baselines and Evaluation Metrics}

For single-frame evaluation, we compare GeoFrameNet with three conventional receivers. \emph{All-Symbol Grid-LS} (Grid-LS) accumulates projection energy over all 64 symbols, selects one valid grid support, estimates one LS coefficient per symbol, and performs adjacent-symbol differential detection. \emph{First-Symbol Sparse Bayesian Learning} (FS-SBL) follows the staged receiver in \cite{Xu2024JointLocalization}, with geometry estimated from the first OFDM-symbol CFR and then fixed for symbol-wise LS estimation and differential detection over the frame. Our off-grid SBL implementation uses at least 10 iterations, a relative-objective tolerance of $10^{-3}$, and at most 200 iterations. \emph{All-Symbol Multiple-Measurement-Vector SBL} (MMV-SBL) jointly processes all 64 symbols with shared support and geometry variables and symbol-specific complex
coefficients for 200 iterations.

For multi-frame localization, both baselines use frozen GeoFrameNet outputs. \emph{Position Averaging} computes $\widehat{\mathbf p}_{\rm PA}=F^{-1}\sum_{f=1}^{F}\widehat{\mathbf p}_f$. \emph{Candidate-Logit Averaging} averages geometry logits across frames, applies the valid-candidate mask and top-16 softmax, and combines the resulting posterior weights with the corresponding candidate coordinates averaged after frame-wise refinement. Neither baseline introduces trainable cross-frame parameters.

For $U$ position estimates, the localization metrics are
\begin{equation}
	\begin{aligned}
		e_{{\rm loc},u}
		&=
		\|\widehat{\mathbf p}_u-\mathbf p_u\|_2,
		\\
		{\rm RMSE}
		&=
		\sqrt{
			\frac{1}{U}
			\sum_{u=1}^{U}e_{{\rm loc},u}^2
		},
		\\
		P_{\rm out}(r_0)
		&=
		\Pr(e_{\rm loc}>r_0),
		\qquad
		r_0=5~\mathrm{m}.
	\end{aligned}
	\label{eq:localization_metrics}
\end{equation}
Here, $U=300$ at each SNR, corresponding to frames for single-frame evaluation and groups for multi-frame evaluation at a fixed $F$. The outage probability is estimated by the fraction of localization errors exceeding $r_0$.

For cross-frame summaries over $\Omega_{\rm SNR}$, we report the geometric mean of the per-SNR RMSE values,
\[
{\rm RMSE}_{\rm GM}(F)
=
\exp\!\left(
\frac{1}{|\Omega_{\rm SNR}|}
\sum_{\omega\in\Omega_{\rm SNR}}
\log {\rm RMSE}_{\omega}(F)
\right),
\]
and the arithmetic mean of the corresponding 5-m outage
probabilities, assigning equal weight to each SNR point.

Statistical comparisons use 10,000 paired percentile-bootstrap replicates at each SNR. Single-frame comparisons resample the same frame indices with replacement for both receivers; multi-frame comparisons resample the same underlying eight-frame-group indices. Individual bits are not resampled, and each BER replicate is computed from the paired frame-wise error counts.

For $\Delta_{\rm BER}={\rm BER}_{\rm GeoFrameNet}-{\rm BER}_{\rm baseline}$, a difference is statistically resolved when its 95\% confidence interval excludes zero. A localization gain or degradation is identified when the proposed-method-to-baseline RMSE-ratio interval lies entirely below or above one, respectively. These pointwise intervals describe test-sample variability for the fixed evaluated receivers, not variability across independent training runs or simultaneous confidence statements across SNRs. The aggregate cross-frame metrics are descriptive summaries.

\subsection{Single-Frame Performance}
\label{sec:sf_performance}

\subsubsection{Data Detection}

\begin{figure}[t]
	\centering
	\includegraphics[width=0.8\columnwidth]{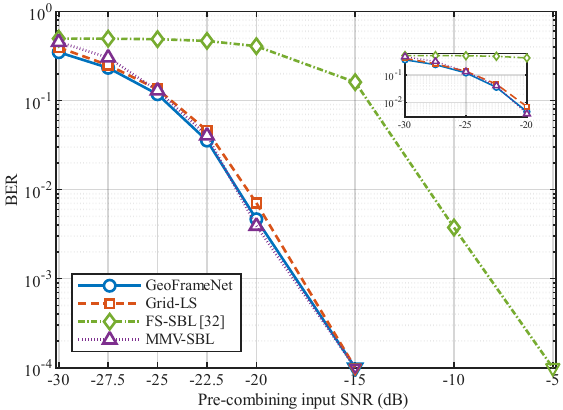}
	\caption{Single-frame BER versus pre-combining input SNR.
		The inset enlarges the weak-to-transition region.
		Values below $10^{-4}$ are clipped; the GeoFrameNet,
		Grid-LS, and MMV-SBL curves terminate at their clipped
		$-15$-dB endpoints.}
	\label{fig:sf_ber}
\end{figure}

Fig.~\ref{fig:sf_ber} shows that GeoFrameNet's detection gains are concentrated in the weak-observation regime. At $-30$~dB, its BER is 0.348995, compared with 0.394868 for Grid-LS and 0.454339 for MMV-SBL, giving a 23.2\% relative reduction over MMV-SBL. At the four weakest evaluated SNR points from $-30$ to $-22.5$~dB, GeoFrameNet reduces both BER and localization RMSE relative to MMV-SBL. This pattern is consistent with using full-frame geometry evidence to weight multiple candidate-conditioned differential logits rather than conditioning detection on one geometry estimate. The multi-hypothesis and gradient-coupling controls in Fig.~\ref{fig:ablation} examine the roles of these design choices.

At $-20$~dB, the BERs of GeoFrameNet, Grid-LS, and MMV-SBL are 0.004656, 0.007090, and 0.003915, respectively. Although MMV-SBL has the lower empirical BER at this point, the paired 95\% confidence interval for the GeoFrameNet-minus-MMV-SBL BER difference is $[-5.82\times10^{-4},\,2.22\times10^{-3}]$. Since this interval includes zero, the observed crossover is not statistically resolved. GeoFrameNet nevertheless retains a statistically resolved BER advantage over Grid-LS at this SNR.

At  $-15$~dB, $-10$ and $-5$~dB, all three receivers record zero observed errors. GeoFrameNet, Grid-LS, and MMV-SBL therefore reach a near-zero-error regime on this finite test set. These observations do not establish zero true BER or BER equivalence.

\subsubsection{Continuous Localization}

\begin{figure}[t]
	\centering
	\includegraphics[width=0.8\columnwidth]{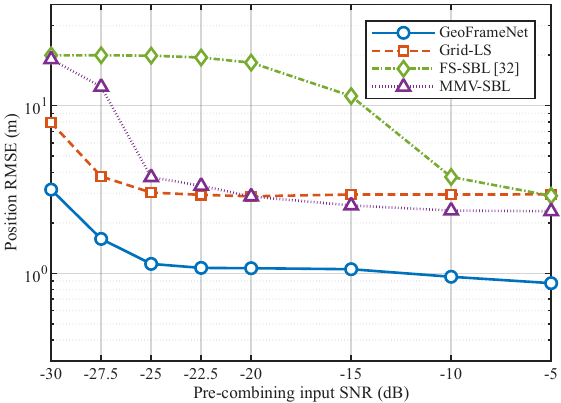}
	\caption{Single-frame two-dimensional localization RMSE
		versus pre-combining input SNR; the vertical axis is
		logarithmic.}
	\label{fig:sf_rmse}
\end{figure}

Fig.~\ref{fig:sf_rmse} shows that GeoFrameNet achieves lower localization RMSE than MMV-SBL at all eight evaluated SNR points. At $-30$~dB, GeoFrameNet achieves an RMSE of 3.1573~m, compared with
7.9180~m for Grid-LS and 18.8859~m for MMV-SBL. The reduction relative to MMV-SBL is 83.3\%. At $-22.5$~dB, the corresponding GeoFrameNet and MMV-SBL values are 1.0771 and 3.3255~m. At every evaluated point, the 95\% interval of the GeoFrameNet-to-MMV-SBL RMSE ratio lies entirely below one.

Because MMV-SBL uses the same number of OFDM symbols, the observed gain is not explained by access to more symbols. GeoFrameNet retains candidate-level weights through continuous position estimation and refines candidate coordinates before combining them. The hard-top-1 and no-offset interventions in Fig.~\ref{fig:ablation} support the roles of these choices within the selected receiver. At stronger tested SNRs, the localization gap remains even where the finite BER test no longer resolves a ranking, showing that near-zero observed BER does not imply comparable continuous position accuracy.

\subsubsection{Localization Error Tail}

\begin{figure}[t]
	\centering
	\includegraphics[width=0.8\columnwidth]{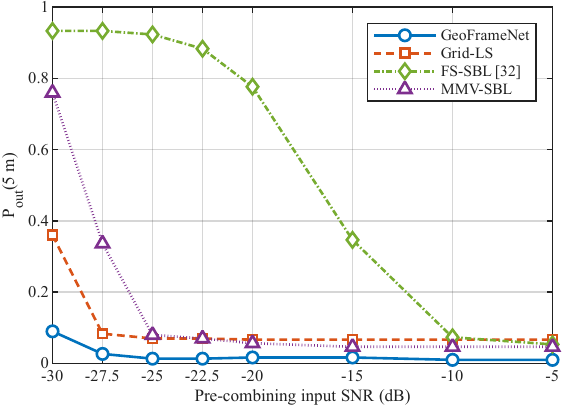}
	\caption{Single-frame localization outage probability
		$\Pr(e_{\rm loc}>5~\mathrm{m})$ versus pre-combining
		input SNR.}
	\label{fig:sf_outage}
\end{figure}

Fig.~\ref{fig:sf_outage} shows that the RMSE improvement is accompanied by fewer large localization errors. At $-30$~dB, the 5-m outage probability is 0.0900 for GeoFrameNet, versus 0.3600 for Grid-LS, 0.9333 for FS-SBL, and 0.7600 for MMV-SBL. At $-27.5$~dB, GeoFrameNet and MMV-SBL yield 0.0267 and 0.3367, respectively. These results show fewer localization errors exceeding 5~m at the reported weak-SNR points, complementing the RMSE gains with a threshold-based measure of localization reliability.

\subsection{Mechanism Ablation}
\label{sec:mechanism_ablation}

\begin{figure*}[t]
	\centering
	\includegraphics[width=0.8\textwidth]{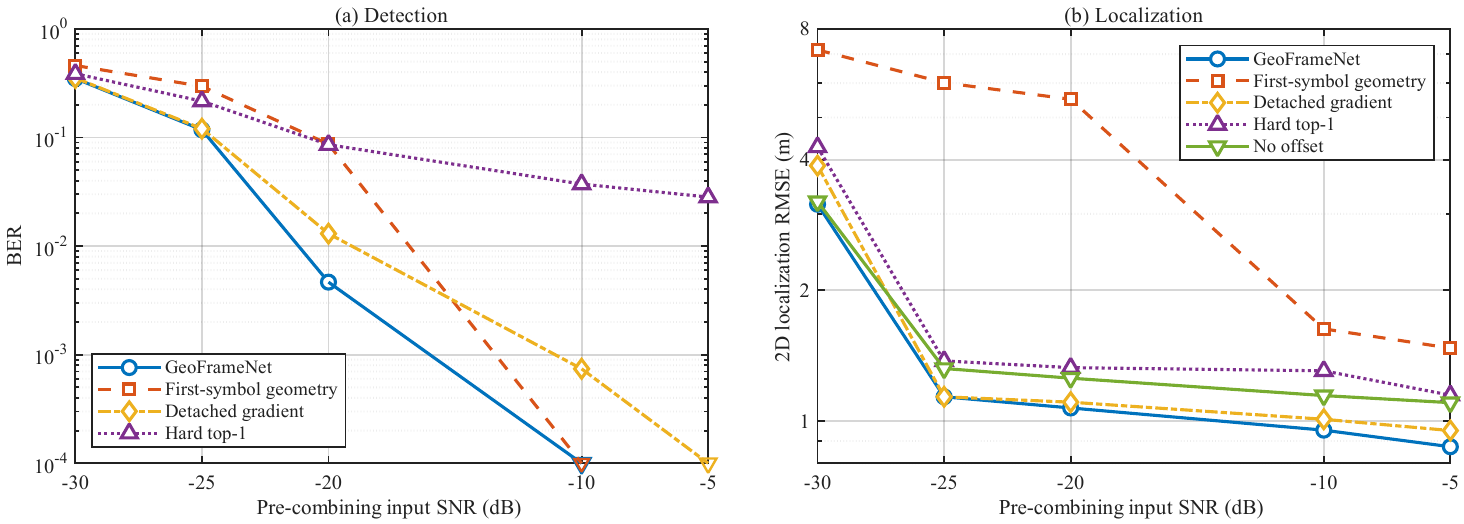}
	\caption{Mechanism ablation of GeoFrameNet for
		(a) data detection and (b) localization. The first-symbol and detached-gradient variants are separately trained controls; hard top-1 and no offset are inference-time interventions on the selected GeoFrameNet model. The no-offset BER curve is omitted because its bit decisions are identical to those of GeoFrameNet. In (a), BER values below $10^{-4}$ are shown at the display floor.} 
	\label{fig:ablation}
\end{figure*}

Fig.~\ref{fig:ablation} examines four receiver mechanisms. At $-20$~dB, the first-symbol geometry control increases BER from 0.0047 to 0.0868 and RMSE from 1.073 to 5.509~m. Hard top-1 inference gives a BER of 0.0858 and an RMSE of 1.328~m even though full-frame geometry inference is retained. The former comparison supports exploiting repeated geometric observations; the latter shows that, for the selected trained receiver, full-frame aggregation alone does not recover the performance of multi-hypothesis task inference. Thus, strengthening geometry evidence and retaining candidate alternatives serve complementary roles.

The detached-gradient control retains full-frame aggregation but blocks the bit-loss gradient through the posterior-derived weights. At $-20$~dB, BER increases to 0.0130, approximately 2.8 times the GeoFrameNet value, whereas RMSE changes only modestly to 1.106~m. The larger relative effect on BER supports using communication supervision to shape detection-relevant geometry scores. At this operating point, the benefit is primarily reflected in detection rather than a comparable improvement in localization accuracy.

Removing continuous offsets leaves the bit decisions unchanged but increases RMSE to 1.256~m at $-20$~dB. Since the detector does not use the refined Cartesian coordinates, this intervention isolates the localization role of continuous refinement. The hard-top-1 and no-offset results assess component use by the selected trained model, not the best attainable performance of separately retrained alternatives.

\subsection{Cross-Frame Localization Performance}

\begin{figure}[t]
	\centering
	\includegraphics[width=0.8\columnwidth]{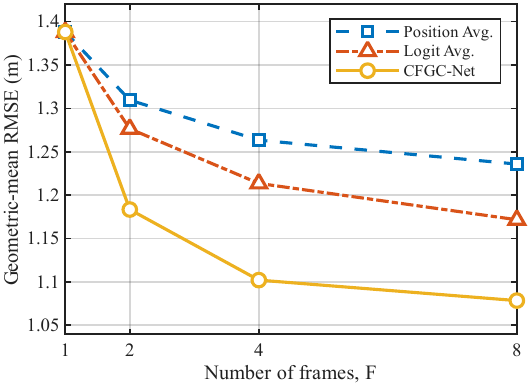}
	\caption{Geometric-mean localization RMSE versus frame
		count across the eight primary SNR points. All methods
		share the same $F=1$ GeoFrameNet reference.}
	\label{fig:mf_rmse}
\end{figure}

\begin{figure}[t]
	\centering
	\includegraphics[width=0.8\columnwidth]{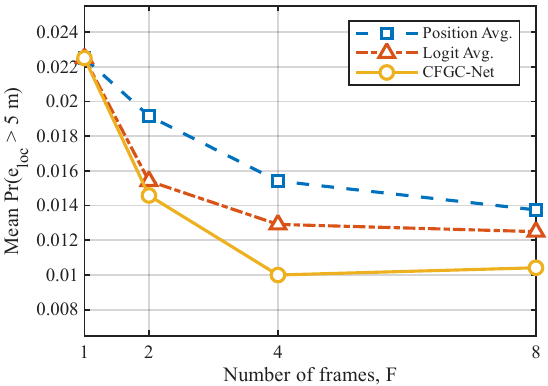}
	\caption{Mean localization outage probability versus
		frame count across the eight primary SNR points.
		All methods share the same $F=1$ GeoFrameNet reference.}
	\label{fig:mf_outage}
\end{figure}

Figs.~\ref{fig:mf_rmse} and~\ref{fig:mf_outage} compare the fusion strategies on the separate multi-frame test groups. Their common $F=1$ reference is frozen GeoFrameNet, with geometric-mean RMSE 1.3880~m and mean outage 0.0225. For $F=2,4,8$, CFGC-Net achieves geometric-mean RMSEs of 1.1834, 1.1021, and 1.0784~m, respectively, with corresponding mean outage probabilities of 0.0146, 0.0100, and 0.0104.

CFGC-Net attains the lowest geometric-mean RMSE and mean outage among the evaluated strategies at each $F\in\{2,4,8\}$. At $F=4$, its geometric-mean RMSE is 12.8\% lower than that of Position Averaging and 9.2\% lower than that of Candidate-Logit Averaging. These comparisons use the same frozen single-frame receiver and the same observations at a fixed frame count. Position Averaging fuses final Cartesian estimates, while Candidate-Logit Averaging fuses geometry logits and frame-wise refined candidate coordinates. CFGC-Net instead uses averaged candidate features for both score correction and frozen continuous refinement. The measured gain therefore reflects the combined score and coordinate construction, not the residual score correction in isolation.

The aggregate ordering does not imply uniformly significant improvements at individual SNRs. Among the $3\times8=24$ frame-count/SNR comparisons with each baseline, the pointwise RMSE-ratio intervals identify six gains over Position Averaging and two over Candidate-Logit Averaging, with no significant degradations. The increase in CFGC-Net's mean outage from 0.0100 at $F=4$ to 0.0104 at $F=8$ shows that this empirical tail metric is not strictly monotonic. Frame-specific bit logits and decisions remain unchanged by construction, so cross-frame fusion does not alter frame-wise BER.

\subsection{Runtime and Model Size}
\label{sec:runtime_complexity}

\begin{table}[t]
	\caption{Measured Processing Time}
	\label{tab:method_complexity}
	\centering
	\footnotesize
	\setlength{\tabcolsep}{5.5pt}
	\begin{tabular}{l c}
		\hline
		Method & Median processing time \\
		\hline
		Grid-LS
		& 0.372 ms/frame \\
		
		FS-SBL~\cite{Xu2024JointLocalization}
		& 2.224 s/frame \\
		
		MMV-SBL
		& 16.455 s/frame \\
		
		GeoFrameNet
		& 1.199 ms/frame \\
		
		CFGC-Net
		& \shortstack{5.177 / 9.073 / 16.924 ms/group\\
			($F=2/4/8$)} \\
		\hline
	\end{tabular}
\end{table}

Table~\ref{tab:method_complexity} reports RTX~5080 implementation-level processing times. GeoFrameNet and Grid-LS use single-precision arithmetic and batches of 50; FS-SBL and MMV-SBL use double-precision complex arithmetic and process one frame at a time. Neural timings include explicit device synchronization. Single-frame entries are the medians of the eight per-SNR medians. CFGC-Net group timings include the $F$ frozen GeoFrameNet passes and subsequent consensus processing.

Grid-LS remains the fastest evaluated implementation. GeoFrameNet requires 1.199~ms/frame and uses 407,812 active inference parameters. CFGC-Net reuses these parameters and adds 24,770 trainable consensus parameters, giving 432,582 active parameters in total. The proposed receiver therefore trades additional processing relative to Grid-LS for the reported localization and weak-observation detection gains. Differences in numerical precision, batching, and implementation prevent interpreting the measured timing gaps as an implementation-independent complexity comparison.

\section{Conclusion}
\label{sec:conclusion}

This paper developed two complementary designs for ambient backscatter reception: GeoFrameNet for single-frame joint continuous localization and data detection, and CFGC-Net for cross-frame localization. GeoFrameNet combines sign-robust full-frame evidence with a shared geometry posterior to avoid a fixed intermediate geometry estimate. Its ablations support complementary roles for evidence aggregation and multi-hypothesis inference, while communication-aware scoring and continuous refinement have task-dependent effects. CFGC-Net extends this candidate-level processing by fusing frozen GeoFrameNet logits and features across frames from a fixed BD while preserving frame-wise detection outputs.

Under the evaluated conditions, GeoFrameNet achieves lower localization RMSE than MMV-SBL at all eight test SNR points, with simultaneous BER and RMSE reductions from $-30$ to $-22.5$~dB. CFGC-Net achieves the lowest aggregate localization RMSE among the evaluated fusion strategies for $F\in\{2,4,8\}$. These findings are limited to the considered single-BD model with a dominant backscatter path and idealized environmental suppression. Richer propagation, imperfect suppression, concurrent multiple-BD reception, and over-the-air validation remain directions for further study.

\end{document}